\pdfoutput=1

\documentclass[preprint,5p,numbers,sort&compress]{elsarticle}

\usepackage{graphicx}
\usepackage{amssymb,amsmath}
\usepackage{natbib}
\usepackage{hyperref}
\usepackage{multicol}
\usepackage{multirow}
 
\hypersetup{pdftitle = The title of my PDF, pdfauthor = My name, pdfsubject= The subject, pdfkeywords = keyword1 keyword2 keyword3}
\hypersetup{colorlinks = true, linkcolor = blue, anchorcolor = red, citecolor = blue, filecolor = red, pagecolor = red, urlcolor = blue}

\begin{document}

\begin{frontmatter}
\title{Orbital dynamics in the photogravitational restricted four-body problem: Lagrange configuration}

\author[jeo]{J. E. Osorio-Vargas}

\author[fld]{F. L. Dubeibe\corref{cor1}}
\ead{fldubeibem@unal.edu.co}

\author[jeo]{Guillermo A. Gonz\'alez}

\cortext[cor1]{Corresponding author}

\address[jeo]{Escuela de F\'{i}sica, Universidad Industrial de Santander, A.A. 678, Bucaramanga, Colombia}

\address[fld]{Facultad de Ciencias Humanas y de la Educaci\'{o}n,
Universidad de los Llanos, Villavicencio, Colombia}

\begin{abstract}
We study the effect of the radiation parameter in the location, stability and orbital dynamics in the Lagrange configuration of the restricted four-body problem when one of the primaries is a radiating body. The equations of motion for the test particle are derived by assuming that the primaries revolve in the same plane with uniform angular velocity, and regardless of their mass distribution, they will always lie at the vertices of an equilateral triangle. The insertion of the radiation factor in the restricted four-body problem, let us model more realistically the dynamics of a test particle orbiting an astrophysical system with an active star. The dynamical mechanisms responsible for the smoothening on the basin structures of the configuration space is related to the decrease in the total number of fixed points with increasing values of the radiation parameter. In our model of the Sun-Jupiter-Trojan Asteroid system, it is found that despite the repulsive character of the solar radiation pressure, there exist two stable libration points roughly located at the position of $L_4$ and $L_5$ in the Sun-Jupiter system.
\end{abstract}


\end{frontmatter}

\section{Introduction}
\label{intro}
The circular restricted three-body problem is one of the most iconic problems in celestial mechanics and was firstly considered by Euler in the context of his lunar theories. This model was refined to a greater extent by the works of Jacobi, Levi-Civita, Birkhoff, Delaunay, and Hill \cite{W14}. Nowadays its theoretical framework constitutes the basis of most of the lunar and planetary theories used in astronautics \cite{B68}. In this model, two massive bodies (primaries) rotate about their barycenter in circular or elliptic trajectories, while the third body moves under the gravitational attraction of the primaries without perturbing their motion. Until today, the study of the dynamics in the restricted three-body problem is still an active field of research (see e.g. \cite{AFKP12,ZD18,DLG17,Z15A,Z15B}).

The planar restricted four-body problem (henceforth PR4BP) is commonly referred to as the restricted (3+1)-body problem. Here, the fourth body has no gravitational effect on the others, such that can be treated as a system composed of a test particle in the presence of three primaries. A group of solutions for (3+1)-body problem refers to the central configurations of the three-body problem, which include the straight-line equilibrium configuration (Euler configuration) and the equilateral triangle configuration (Lagrange configuration) \cite{L06, H16}. The straight-line equilibrium configuration of the PR4BP can be formally derived by introducing an additional primary to the restricted three-body problem in the position of the barycenter, such that the primaries are always in syzygy, while for the Lagrange configuration the primaries are located at the vertices of an equilateral triangle (see e.g. \cite{M81}). 

The Euler and Lagrange configurations of the PR4BP have been widely studied in the literature, ranging from the calculation of the equilibrium points and their stability \cite{MP08,M81,BP11B} to the computation of families of periodic orbits \cite{BP11A, BD13, S78, ARB15, BGD13}, or the study of the orbital dynamics of escape and collision in these systems \cite{ZE16,ML98}. Over the years, several modifications to the basic  Euler and Lagrange configurations have been investigated to understand the influence and effects of different parameters  in realistic celestial systems. Papadouris and Papadakis studied, under the assumption of two equal masses, the existence, location, and stability of the equilibrium points in the photogravitational version of the Lagrange configuration \cite{PP13}, and the periodic solutions for this system \cite{PP14}. Singh and Vincent conducted a similar study but considering all primaries as radiation sources \cite{SV16}. Concerning the shape of the primaries, Asique et al. studied the location of the libration points and their stability in the photogravitational PR4BP when one of the primaries is an oblate/prolate spheroid \cite{APHS16}, while in Chand et.al. \cite{CUHS15}, the stability is studied when the third primary is an oblate spheroid.

Since the incorporation of radiation pressure on the primary bodies affects the existence and stability of the equilibrium points, it is important to consider the effect of the radiation on the orbital dynamics of the system. In the present paper, we shall study the four-body problem in the Lagrangian configuration by considering the primary body $m_1$ as the radiation source. Here, we extend the works by Baltagiannis \& Papadakis \cite{BP11B}, Zotos \cite{ZE16}, and Papadouris \& Papadakis  \cite{PP13}, by considering not only three different combinations of mass for the primary bodies (three equal masses, two equal masses, and three different masses), but also the radiation pressure. Moreover, taking into account that in our Solar system the Sun, Jupiter and the Trojan asteroids form an equilateral triangle configuration, we will consider the characteristic values of this system for the case of  three different masses. 

The structure of the paper is as follows: In section \ref{mod}, we give a brief formulation of the problem and then present the equations of motion. In section \ref{libration}, we derive general equations to determine the location and stability of the libration points and also describe the parametric evolution of these points as a function of the radiation factor. In section  \ref{results}, we outline the numerical criteria used for the classification of orbits and show the parametric evolution of the orbital structure in the photogravitational Lagrange configuration of the PR4BP, for the three combinations of mass under consideration. We end with the main conclusions of this research in section \ref{conc}.

\section{Formulation of the problem and equations of motion}
\label{mod}

The photogravitational Lagrangian configuration of the restricted four-body problem describes the motion of a test particle $m$ under the gravitational field of three massive bodies, $m_1$, $m_2$, and $m_3$, with at least one of them being a radiation source. The primaries revolve in the same plane with uniform angular velocity, and regardless of their mass distribution, they will always lie at the vertices of an equilateral triangle. 

\begin{figure}[h]
\centering
\resizebox{\hsize}{!}{\includegraphics[scale = 1]{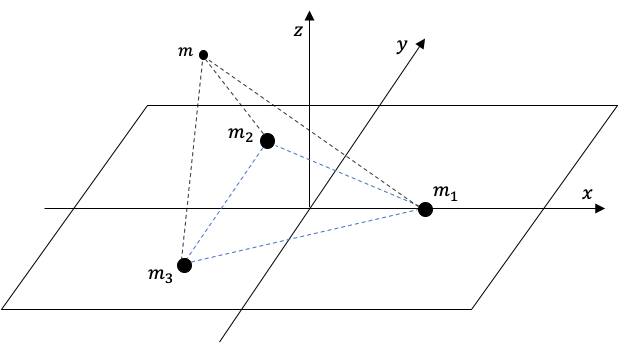}}
\caption{Schematic representation of the Lagrangian configuration for the restricted four-body problem.} 
\label{fig_1}
\end{figure}

By using canonical units, the sum of masses of the primaries, as well as the distance between them, the angular velocity, and the gravitational constant are normalized to one. With the barycenter shifted to the origin of the synodic frame of reference, the positions of the primaries $(x_i, y_i)$ can be calculated as a function of the masses, yielding four possible combinations. Here, we choose the arrangement in which the primary $m_1$ is situated on the positive $x-$axis and $m_2$ lies on the second quadrant (see Fig. \ref{fig_1}), such that the explicit expressions for the positions are given by (see e.g.  \cite{M00})

\begin{eqnarray}
x_1 &=& \sqrt{m_2^2 + m_2 \, m_3 + m_3^2} \, ,\label{coord1}\\
x_2 &=&- \dfrac{m_3 \, (m_2 - m_3) + m_1 \, (2 \, m_2 + m_3)}{2 \, \sqrt{m_2^2 + m_2 \, m_3 + m_3^2}} \, ,\\
x_3 &=& -\dfrac{m_2 \, (m_3 - m_2) + m_1 \, (m_2 + 2 \, m_3)}{2 \, \sqrt{m_2^2 + m_2 \, m_3 + m_3^2}} \, ,\\
y_1 &=& 0 \, ,\\
y_2 &=& \dfrac{\sqrt{3}}{2} \, \dfrac{m_3}{\sqrt{m_2^2 + m_2 \, m_3 + m_3^2}} \, ,\\
y_3 &=& -\dfrac{\sqrt{3}}{2} \, \dfrac{m_2}{\sqrt{m_2^2 + m_2 \, m_3 + m_3^2}} \, .
\label{coord6}
\end{eqnarray}

With the previous considerations, the equations of motion for the photogravitational Lagrange configuration of the PR4BP in the synodic frame can be written as
\begin{eqnarray}
\ddot{x} - 2 \dot{y} &=& x - \sum_{i=1}^{3} \dfrac{(1 - \beta_i) m_i (x - x_i)}{r_i^3},\label{eqs_motion1}\\
\ddot{y} + 2 \dot{x} &=& y - \sum_{i=1}^{3} \dfrac{(1 - \beta_i) m_i (y - y_i)}{r_i^3}, \label{eqs_motion2}
\end{eqnarray}
with $r_i = \sqrt{(x - x_i)^2 + (y - y_i)^2}$, and $i = 1,2,3$. The radiation factor is by definition a dimensionless number in the interval $\beta \in [0,1]$ \cite{BLS79},  representing the quotient between the solar radiation pressure force $F_{r}$ and the gravitational attraction force $F_g$ \cite{S80}. Hereafter, we set $\beta_2 = \beta_3 = 0$, i.e., only the primary $m_1$ is a radiating body.

In compact form, the equations of motion for the test particle in the synodical frame read as
\begin{eqnarray}
\ddot{x} - 2 \, \dot{y} &=& \dfrac{\partial U}{\partial x}, \label{eqm1}\\
\ddot{y} + 2 \, \dot{x} &=& \dfrac{\partial U}{\partial y},\label{eqm2}
\end{eqnarray}
where
\begin{eqnarray}
U = \dfrac{1}{2} (x^2 + y^2) + (1-\beta)\dfrac{m_1}{r_1} + \dfrac{m_2}{r_2} + \dfrac{m_3}{r_3}.
\end{eqnarray}

The dynamical system (\ref{eqm1}-\ref{eqm2}) admits the well-known integral of motion 
\begin{eqnarray}
C = 2 U - (\dot{x}^2 + \dot{y}^2),
\end{eqnarray}
which restricts the motion of the test particle to the region $C \leq 2 U$.

It is important to notice that throughout the paper we shall assume that the primaries $m_1$, $m_2$, and $m_3$ hold in an equilateral triangle despite the extra forces. This assumption is based on the fact that the effects of radiation pressure of $m_1$ on $m_2$ and $m_3$ are very small and can be neglected.

\section{Libration points and linear stability}
\label{libration}

As noted in the introduction section, during the last decade some authors have focused on the study of the equilibrium points and their stability in the Lagrangian configuration for the PR4BP \cite{BP11B} and the photogravitational PR4BP \cite{PP13}, by considering the particular case of two equal masses for the primaries. In this section, we attempt to show how, besides the radiation factor $\beta$, the different combinations of mass ($m_1=m_2=m_3$, $m_1 \neq m_2=m_3$, and $m_1 \neq m_2 \neq m_3$) affect the location and stability of the libration points.

The libration points can be calculated by imposing the conditions: $\dot{x} = \dot{y} = \ddot{x} = \ddot{y} = 0$, while the stability can be determined by linearizing the equations of motion Eqs. (\ref{eqm1}-\ref{eqm2}) about a fixed point ($x^*, y^*$). This process leads to the linearized state-space equation $\vec{\dot{\eta}}=\mathbb{A} \vec{\eta}$, where $\vec{\eta}=(x, y, \eta_1, \eta_2)^{T}$, with $(\eta_1, \eta_2)=(\dot{x}, \dot{y})$, whose coefficient matrix reads as
\begin{equation}
\mathbb{A} =
\left(
\begin{array}{cccc}
0 & 0 & 1 & 0\\
0 & 0 & 0 & 1\\
A_{11} & A_{12} & A_{13} & A_{14}\\
A_{21} & A_{22} & A_{23} & A_{24}
\end{array}
\right),
\end{equation}
with
\begin{eqnarray*}
A_{11} &=& 1 + \dfrac{m_1 \left[2(x - x_1)^2 - (y - y_1)^2 \right]}{\left[(x - x_1)^2 + (y - y_1)^2\right]^{5/2}}(1 - \beta)\\ &+& \sum_{i=2}^{3} \dfrac{m_i \left[2(x - x_i)^2 - (y - y_i)^2 \right]}{\left[(x - x_i)^2 + (y - y_i)^2\right]^{5/2}},\\
A_{12} &=& \dfrac{3 m_1 (x - x_1)(y - y_1)}{\left[(x - x_1)^2 + (y - y_1)^2\right]^{5/2}}(1 - \beta)\\ &+& 3 \sum_{i=2}^{3} \dfrac{m_i(x - x_i)(y - y_i)}{\left[(x - x_i)^2 + (y - y_i)^2\right]^{5/2}},\\
A_{13} &=& 0,\\
A_{14} &=& 2,\\
A_{21} &=& A_{12},\\
A_{22} &=& 1 - \dfrac{m_1 \left[(x - x_1)^2 - 2(y - y_1)^2 \right]}{\left[(x - x_1)^2 + (y - y_1)^2\right]^{5/2}}(1 - \beta)\\ &-& \sum_{i=2}^{3} \dfrac{m_i \left[(x - x_i)^2 - 2(y - y_i)^2 \right]}{\left[(x - x_i)^2 + (y - y_i)^2\right]^{5/2}},\\
A_{23} &=& -2,\\
A_{24} &=&  0.
\end{eqnarray*}

Accordingly, the characteristic polynomial is given by 
\begin{equation}
P(\lambda) =\lambda^4 + \lambda^2 (4  - A_{11} - A_{22}) + A_{11}\,A_{22} - A_{12}^2.
\label{eq:charac_eq}
\end{equation}

The libration points are said to be stable if the roots of $P(\lambda)=0$, evaluated at $(x^{*},y^{*})$ are complex with negative real parts or pure imaginary. In all the other cases, the libration points are unstable.

\subsection{Case 1: $m_1 = m_2 = m_3$}\label{3.1}

We start by considering the case in which the three primary bodies have the same mass, $m_1=m_2=m_3=1/3$. From Eqs. (\ref{coord1}-\ref{coord6}), the location of the primaries ($x_1,y_1$), ($x_2,y_2$), and ($x_3,y_3$), are respectively
\begin{equation*}
\left(\frac{1}{\sqrt{3}},0\right),   \left(-\frac{1}{2\sqrt{3}}, \frac{1}{2}\right), \quad {\rm and} \quad  \left(-\frac{1}{2\sqrt{3}}, -\frac{1}{2}\right). 
\end{equation*}

\begin{figure}[h]
\centering
\resizebox{\hsize}{!}{\includegraphics[width = \columnwidth]{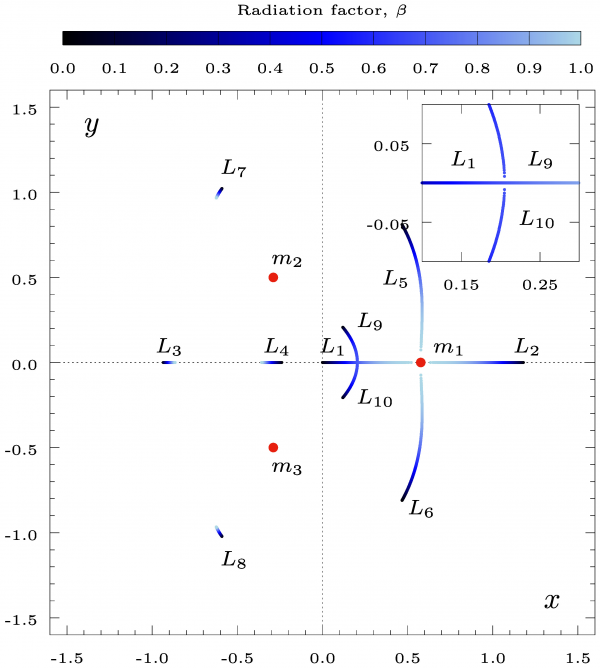}}
\caption{(Color online). Parametric evolution of the libration points for $\beta$ increasing from 0 to 1. 
As indicated in the color palette, darker blue indicates a lower value of radiation factor while lighter blue indicates a higher value of radiation factor. The primaries are represented by red dots.  $L_9$ and $L_{10}$ disappear for $\beta\approx 0.690$ as depicted in the enlargement (inset).} 
\label{fig_2}
\end{figure}

In Fig. \ref{fig_2}, we illustrate the evolution of the equilibrium points for a radiation factor $\beta$ varying from 0 to 1. Through the present section, we shall use a color palette to represent the variation of the radiation factor $\beta$, where darker blue indicates a lower value of radiation factor ($\beta\rightarrow 0$), and lighter blue indicates a higher value of radiation factor ($\beta\rightarrow 1$). The positions of the primaries are represented by red dots. Here, it is observed that as $\beta$ increases, $L_1$, $L_2$, $L_5$, and $L_6$, tend to $m_1$, but they disappear before reaching its position at $\beta\approx 0.999$.  Also, it can be seen that the Lagrangian points $L_9$ and $L_{10}$ tend to the same point along the $x-$axis and completely disappear for $\beta\approx 0.690$. On the other hand, $L_3$, $L_4$, $L_7$, and $L_8$, show slight displacements from their initial positions.

Regarding the stability of the fixed points, it is found that in accordance with the results of Ref. \cite{BP11B}, for $\beta = 0$, all the equilibrium points are unstable. Furthermore, for values of the radiation parameter greater than zero, $\beta \in (0,1]$, the stability of the librations remains unaltered, since the corresponding eigenvalues $\lambda_{1,2,3,4}$, have the form: {\it{(i)}} $\lambda_{1,2,3,4} = \pm a \pm ib$ for $L_1$, $L_3$, $L_5$ and $L_6$ and {\it{(ii)}} $\lambda_{1,2} = \pm \, ib$ and $\lambda_{3,4} = \pm \, a$ for $L_2$, $L_4$, $L_7$, $L_8$, $L_9$ and $L_{10}$. Therefore, we may conclude that if the primary bodies are equal mass, all the equilibrium points of the photogravitational Lagrange configuration of the PR4BP are unstable regardless of the value of $\beta$. We refer the reader to Table \ref{table_1} for additional information about the existence and stability of the libration points. 

\begin{table}[h]
\caption{Existence and stability of equilibrium points with the variation of the radiation factor $\beta$, for $m_1= m_2 = m_3=1/3$. \label{table_1}}
\vspace{0.2cm}
{\begin{tabular}{cccc}
\hline
{\text{Interval}} & {\text{Equilibria}} & {\text{Stable equilibria}} \\ 
[0.1cm]\hline\\
[-0.2cm]
$\beta \in [0.000, 0.690]$ & $L_{1,2,3,4,5,6,7,8,9,10}$ & $-$\\
$\beta \in [0.691, 0.999]$ & $L_{1,2,3,4,5,6,7,8}$ & $-$\\
$\beta = 1.000$ & $L_{3,4,7,8}$ & $-$ 
\\
\hline
\end{tabular}}
\end{table}

\subsection{Case 2: $m_1 \neq m_2 = m_3$}\label{3.2}

As a second case, we consider the situation in which two primaries have the same mass $m_2 = m_3={\mathfrak{m}}$. Consequently, the mass of the first primary is given by $m_1 = 1 - 2  {\mathfrak{m}}$, while the positions of the primaries $(x_1, y_1), (x_2, y_2),$ and $(x_3, y_3)$, according to Eqs. (\ref{coord1}-\ref{coord6}), read as follows
\begin{equation*}
\left(\sqrt{3} {\mathfrak{m}}, 0\right),   \left(\frac{\sqrt{3}}{2}(2 {\mathfrak{m}}-1), \frac{1}{2}\right), \,\, {\rm and} \,\,  \left(\frac{\sqrt{3}}{2}(2 {\mathfrak{m}}-1), -\frac{1}{2}\right). 
\end{equation*}

\begin{figure*}[t!]
\centering
\includegraphics[width = \linewidth]{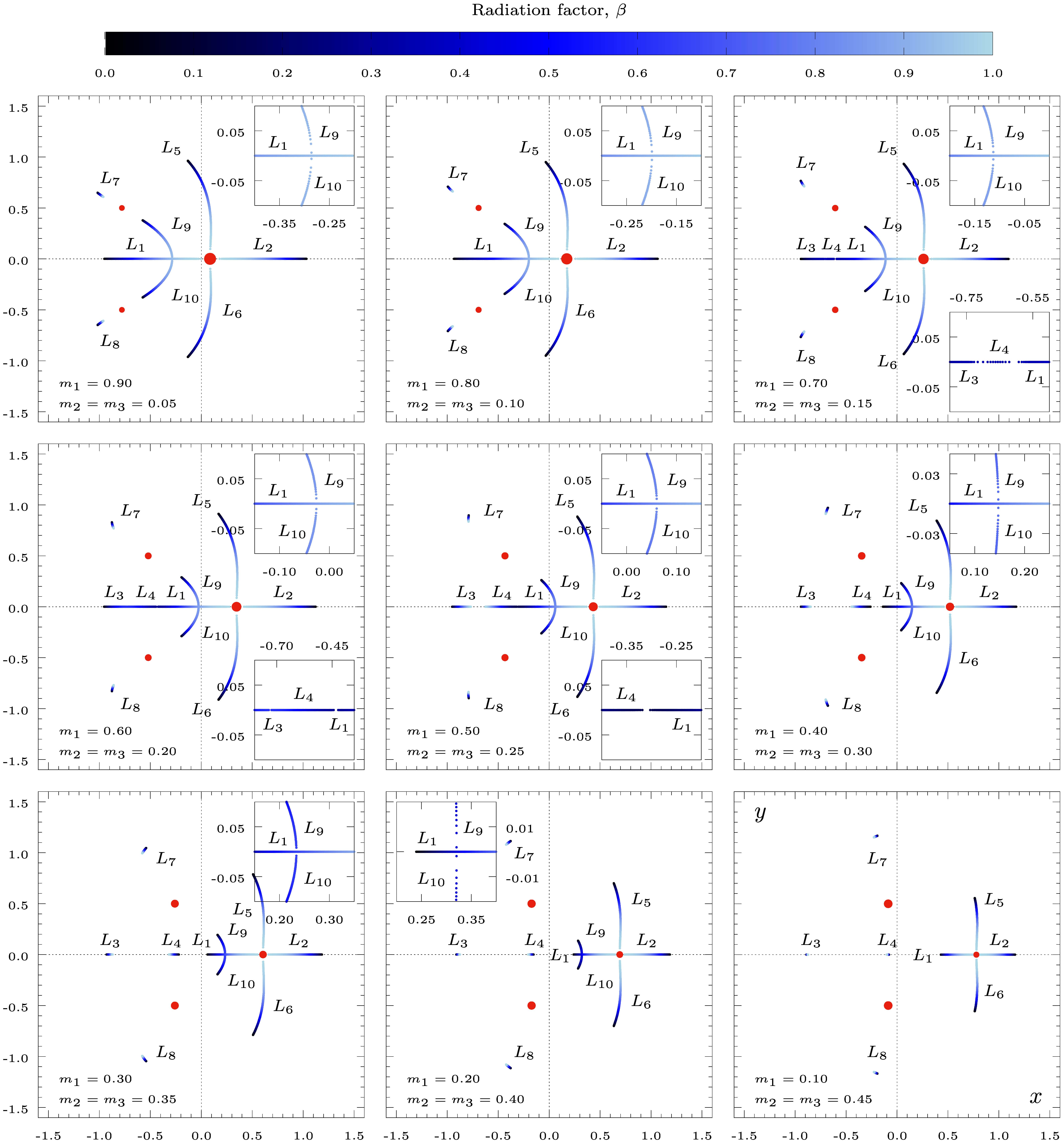}
\caption{(Color online). Parametric evolution of the equilibrium points for $\beta$ increasing from 0 to 1, where each panel represents a different value of ${\mathfrak{m}}=m_2 = m_3$. As indicated in the color palette, darker blue indicates a lower value of radiation factor while lighter blue indicates a higher value of radiation factor. The primaries are represented by red dots. A closer look at the variation of  $L_{1}, L_{9}$, and $L_{10}$ is depicted in the insets.}
\label{fig_3}
\end{figure*}

Therefore, the existence and location of libration points depend on the parameters ${\mathfrak{m}}$ and $\beta$. In Figure \ref{fig_3}, we present the variation in the position of equilibrium points, for a radiation factor $\beta$ increasing from 0 to 1. Each panel represents a different combination of masses for the primaries. In the first two panels, $m_1=0.9$ and $m_1=0.8$, it is observed that 
the collinear points $L_3$ and $L_4$ do not exist. For $0.2\leq m_1\leq 0.7$, ten equilibria exist, while for $m_1=0.1$ the non-collinear points $L_9$ and $L_{10}$ disappear. 

Concerning the evolution of the equilibrium points in terms of the radiation parameter, in Fig. \ref{fig_3} 
it is observed that, for larger values of $\beta$, the equilibria $L_9$ and $L_{10}$  move to the $x-$axis, such that they almost join each other. $L_3$, $L_4$, $L_7$, and $L_8$, exhibit very small displacements with respect to their initial positions, while the libration points $L_1$, $L_2$, $L_5$, and $L_6$, are displaced toward the radiating body, but disappear for $\beta\simeq1$. Additionally, in Table \ref{table_2}, we present the total number of equilibrium points in terms of $\beta$ and ${\mathfrak{m}}$. It can be observed that for larger values of $\beta$ the total number of fixed points tends to be reduced, however, the final number of surviving librations is larger for larger values of ${\mathfrak{m}}$.

For the analysis of the linear stability of the equilibrium solution, we begin considering the case $\beta = 0$. The results can be summarized as follows: for $\mathfrak{m} \in (0, 0.0027]$, $L_1$, $L_5$, and $L_6$ are stable; for $\mathfrak{m} \in [0.0027, 0.0188]$,  only $L_5$ and $L_6$ keeps the stability; and for $\mathfrak{m}>0.0188$ all the equilibria are unstable. These results are in agreement with Ref. \cite{BP11B}. In the case $\beta>0$, Table \ref{table_3} provides a detailed description of the stability of the equilibrium points. By varying $\beta$ in steps of $\Delta \beta = 1 \times 10^{-2}$, we obtain the mass ranges for which $L_1$, $L_5$, and $L_6$ are stable, depending on the radiation factor $\beta$. Also, it can be concluded that the remaining libration points, $L_2, L_3, L_4, L_7, L_8, L_9,$ and $L_{10}$, are always unstable no matter the values set for the radiation factor $\beta$ or the mass for the primaries $\mathfrak{m}$. 

\begin{table}[t!]
\centering
\caption{Existence of equilibrium points with the variation of the radiation factor $\beta$, for different values of $m_2 = m_3 = {\mathfrak{m}}$. \label{table_2}}
\vspace{0.2cm}
{\begin{tabular}{ccc}
\hline
{${\mathfrak{m}}$} & {\text{Interval}} & {\text{Equilibria}} \\ 
[0.1cm]\hline\\
[-0.2cm]
\multirow{3}{*}{$0.05$}& $\beta \in [0.000, 0.934]$ & $L_{1,2,5,6,7,8,9,10}$ \\
& $\beta \in [0.935, 0.999]$ & $L_{1,2,5,6,7,8}$\\
& $\beta = 1.000$ & $L_{7,8}$ \\\hline
\multirow{3}{*}{$0.10$} & $\beta \in [0.000, 0.916]$ & $L_{1,2,5,6,7,8,9,10}$ \\
& $\beta \in [0.917, 0.999]$ & $L_{1,2,5,6,7,8}$ \\ 
& $\beta = 1.000$ & $L_{7,8}$ \\\hline
\multirow{5}{*}{$0.15$} & $\beta \in [0.000, 0.350]$ & $L_{2,3,5,6,7,8,9,10}$ \\
& $\beta \in [0.351, 0.360]$  & $L_{1,2,3,4,5,6,7,8,9,10}$ \\ 
& $\beta \in [0.361, 0.893]$  & $L_{1,2,5,6,7,8,9,10}$ \\
& $\beta \in [0.894, 0.999]$  & $L_{1,2,5,6,7,8}$ \\
& $\beta = 1.000$ & $L_{7,8}$ \\\hline
\multirow{5}{*}{$0.20$} & $\beta \in [0.000, 0.293]$ & $L_{2,3,5,6,7,8,9,10}$ \\
& $\beta \in [0.294, 0.608]$  & $L_{1,2,3,4,5,6,7,8,9,10}$ \\ 
& $\beta \in [0.609, 0.862]$ & $L_{1,2,5,6,7,8,9,10}$ \\
& $\beta \in [0.863, 0.999]$  & $L_{1,2,5,6,7,8}$ \\
& $\beta = 1.000$  & $L_{7,8}$ \\\hline
\multirow{4}{*}{$0.25$} & $\beta \in [0.000, 0.160]$ & $L_{2,3,5,6,7,8,9,10}$ \\
& $\beta \in [0.161, 0.819]$  & $L_{1,2,3,4,5,6,7,8,9,10}$ \\ 
& $\beta \in [0.820, 0.999]$  & $L_{1,2,3,4,5,6,7,8}$ \\
& $\beta = 1.000$ & $L_{3,4,7,8}$ \\\hline
\multirow{3}{*}{$0.30$} & $\beta \in [0.000, 0.755]$ & $L_{1,2,3,4,5,6,7,8,9,10}$ \\
& $\beta \in [0.756, 0.999]$ & $L_{1,2,3,4,5,6,7,8}$ \\ 
& $\beta = 1.000$ & $L_{3,4,7,8}$ \\\hline
\multirow{3}{*}{$0.35$} & $\beta \in [0.000, 0.647]$ & $L_{1,2,3,4,5,6,7,8,9,10}$ \\
& $\beta \in [0.648, 0.999]$ & $L_{1,2,3,4,5,6,7,8}$ \\ 
& $\beta = 1.000$ & $L_{3,4,7,8}$ \\\hline
\multirow{3}{*}{$0.40$} & $\beta \in [0.000, 0.433]$ & $L_{1,2,3,4,5,6,7,8,9,10}$ \\
& $\beta \in [0.434, 0.999]$ & $L_{1,2,3,4,5,6,7,8}$ \\ 
& $\beta = 1.000$ & $L_{3,4,7,8}$ \\\hline
\multirow{2}{*}{$0.45$} & $\beta \in [0.000, 0.999]$ & $L_{1,2,3,4,5,6,7,8}$ \\
& $\beta = 1.000$ & $L_{3,4,7,8}$ \\
\hline
\end{tabular}}
\end{table}

\begin{table}[t!]
\caption{Stability of the equilibrium points with the variation of the radiation factor $\beta$, for $m_2 = m_3 = {\mathfrak{m}}$. \label{table_3}}
\vspace{0.2cm}
{\begin{tabular}{ccc}
\hline
{\text{Radiation interval}} & \text{Mass interval} & {\text{Stable equilibria}} \\ 
[0.1cm]\hline\\
[-0.2cm]
\multirow{2}{*}{$\beta \in [0.00, 0.60]$} & ${\mathfrak{m}} \in (0.000, 0.002]$ & $L_{1,5,6}$ \\
& ${\mathfrak{m}} \in [0.003, 0.018]$ & $L_{5,6}$ \\ \hline
\multirow{2}{*}{$\beta \in [0.61, 0.70]$} & ${\mathfrak{m}} \in (0.000, 0.003]$ & $L_{1,5,6}$ \\
& ${\mathfrak{m}} \in [0.004, 0.018]$ & $L_{5,6}$ \\ \hline
\multirow{2}{*}{$\beta \in [0.71, 0.75]$} & ${\mathfrak{m}} \in (0.000, 0.004]$ & $L_{1,5,6}$ \\
& ${\mathfrak{m}} \in [0.005, 0.018]$ & $L_{5,6}$ \\ \hline
\multirow{2}{*}{$\beta \in [0.76, 0.79]$} & ${\mathfrak{m}} \in (0.000, 0.005]$ & $L_{1,5,6}$ \\
& ${\mathfrak{m}} \in [0.006, 0.018]$ & $L_{5,6}$ \\ \hline
\multirow{2}{*}{$\beta = 0.80$} & ${\mathfrak{m}} \in (0.000, 0.006]$ & $L_{1,5,6}$ \\
& ${\mathfrak{m}} \in [0.007, 0.018]$ & $L_{5,6}$ \\ \hline
\multirow{2}{*}{$\beta = 0.81$} & ${\mathfrak{m}} \in (0.000, 0.006]$ & $L_{1,5,6}$ \\
& ${\mathfrak{m}} \in [0.007, 0.019]$ & $L_{5,6}$ \\ \hline
\multirow{2}{*}{$\beta \in [0.82, 0.83]$} & ${\mathfrak{m}} \in (0.000, 0.007]$ & $L_{1,5,6}$ \\
& ${\mathfrak{m}} \in [0.008, 0.019]$ & $L_{5,6}$ \\ \hline
\multirow{2}{*}{$\beta = 0.84$} & ${\mathfrak{m}} \in (0.000, 0.008]$ & $L_{1,5,6}$ \\
& ${\mathfrak{m}} \in [0.009, 0.019]$ & $L_{5,6}$ \\ \hline
\multirow{2}{*}{$\beta = 0.85$} & ${\mathfrak{m}} \in (0.000, 0.009]$ & $L_{1,5,6}$ \\
& ${\mathfrak{m}} \in [0.010, 0.019]$ & $L_{5,6}$ \\ \hline
\multirow{2}{*}{$\beta = 0.86$} & ${\mathfrak{m}} \in (0.000, 0.010]$ & $L_{1,5,6}$ \\
& ${\mathfrak{m}} \in [0.011, 0.019]$ & $L_{5,6}$ \\ \hline
\multirow{2}{*}{$\beta = 0.87$} & ${\mathfrak{m}} \in (0.000, 0.011]$ & $L_{1,5,6}$ \\
& ${\mathfrak{m}} \in [0.012, 0.019]$ & $L_{5,6}$ \\ \hline
\multirow{2}{*}{$\beta = 0.88$} & ${\mathfrak{m}} \in (0.000, 0.013]$ & $L_{1,5,6}$ \\
& ${\mathfrak{m}} \in [0.014, 0.019]$ & $L_{5,6}$ \\ \hline
\multirow{2}{*}{$\beta = 0.89$} & ${\mathfrak{m}} \in (0.000, 0.015]$ & $L_{1,5,6}$ \\
& ${\mathfrak{m}} \in [0.016, 0.019]$ & $L_{5,6}$ \\ \hline
\multirow{2}{*}{$\beta = 0.90$} & ${\mathfrak{m}} \in (0.000, 0.018]$ & $L_{1,5,6}$ \\
& ${\mathfrak{m}} = 0.019$ & $L_{5,6}$ \\ \hline
\multirow{2}{*}{$\beta = 0.91$} & ${\mathfrak{m}} \in (0.000, 0.020]$ & $L_{1,5,6}$ \\
& ${\mathfrak{m}} \in [0.021, 0.023]$ & $L_{1}$ \\ \hline
\multirow{2}{*}{$\beta = 0.92$} & ${\mathfrak{m}} \in (0.000, 0.020]$ & $L_{1,5,6}$ \\
& ${\mathfrak{m}} \in [0.021, 0.032]$ & $L_{1}$ \\ \hline
\multirow{2}{*}{$\beta = 0.93$} & ${\mathfrak{m}} \in (0.000, 0.020]$ & $L_{1,5,6}$ \\
& ${\mathfrak{m}} \in [0.021, 0.062]$ & $L_{1}$ \\ \hline
\multirow{2}{*}{$\beta = 0.94$} & ${\mathfrak{m}} \in (0.000, 0.020]$ & $L_{1,5,6}$ \\
& ${\mathfrak{m}} \in [0.021, 0.030]$ & $L_{1}$ \\ \hline
\multirow{1}{*}{$\beta \in [0.95, 0.97]$} & ${\mathfrak{m}} \in (0.000, 0.021]$ & $L_{5,6}$ \\ \hline
\multirow{1}{*}{$\beta = 0.98$} & ${\mathfrak{m}} \in (0.000, 0.022]$ & $L_{5,6}$ \\ \hline
\multirow{1}{*}{$\beta = 0.99$} & ${\mathfrak{m}} \in (0.000, 0.023]$ & $L_{5,6}$ \\
\hline
\end{tabular}}
\end{table}

\subsection{Case 3: $m_1 \neq m_2 \neq m_3$}\label{3.3}

Among the infinite set of possibilities for $m_1 \neq m_2 \neq m_3$, we consider the system Sun-Jupiter-Trojan asteroid, because it is of astrophysical interest and forms a natural equilateral triangle configuration. In this configuration, the Sun corresponds to the radiating body $m_1$.  Following Baltagiannis \& Papadakis \cite{BP13}, it is straightforward to compute the normalized masses for the primaries, which are given by  
$m_1 = m_{S} \simeq 0.999046321943$, $m_2 = m_{J} \simeq 0.000953678050$ and $m_3 = m_{A} \simeq 6.99996 \times 10^{-12}$, where $m_{S}$ denotes the normalized mass of the Sun, $m_{J}$ the normalized mass of Jupiter, and $m_{A}$ corresponds to the normalized mass of Hektor, the largest Jupiter trojan. With the previous values of mass and the aid of Eqs. (\ref{coord1}-\ref{coord6}), the positions of the primary bodies can be computed as
\begin{eqnarray*}
(x_1, y_1)&\simeq &(9.53678\times 10^{-4}, 0.0),\\  
(x_2, y_2)&\simeq &(-0.999046, 6.35659\times10^{-9}), \\ 
(x_3, y_3)&\simeq &(-0.499046, -0.866025).
\end{eqnarray*}

Due to the very small mass of the Trojan asteroid in comparison with the other primaries, the problem approaches a three-body problem configuration, i.e, the position of the massive primary $m_1$ nearly coincides with the barycenter, while the primary $m_2$ practically lies on the $x-$axis. However, it should be pointed out that in absence of radiation ($\beta=0$) a non-zero value of mass for $m_3$, allows the existence of three libration points, additional to the five well-known collinear and triangular points of the restricted three-body problem (see \cite{BP13}).

\begin{figure}[ht!]
\centering
\resizebox{\hsize}{!}{\includegraphics[width = \columnwidth]{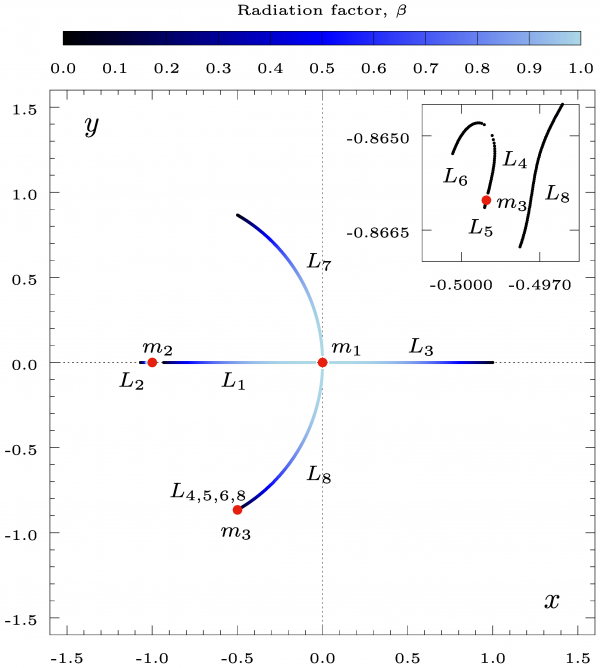}}
\caption{(Color online). Parametric evolution of the libration points for the Sun-Jupiter-Trojan Asteroid system, for $\beta \in \left[0, 1\right]$. As indicated in the color palette, darker blue indicates a lower value of radiation factor while lighter blue indicates a higher value of radiation factor. The primaries are represented by red dots. In the upper right corner, we show a zoom of the parametric evolution around $m_3$.} 
\label{fig_4}
\end{figure}

The parametric evolution of the libration points for the Sun-Jupiter-Trojan Asteroid system, for $\beta \in \left[0, 1\right]$ is presented in Fig. \ref{fig_4}. As indicated in the previous paragraph, for ($\beta = 0$) there exist eight libration points, four of them ($L_4, L_5, L_6$, and $L_8$) nearly coincide with the position of $m_3$. When $\beta>0$, the equilibria $L_4$ and $L_6$ tend to the same point but disappear for $\beta \simeq 0.003$. On the other hand, $L_1$, $L_3$, $L_7$, and $L_8$ move toward the radiating body $m_1$ and vanish for $\beta \simeq 0.999$, while $L_2$ and $L_5$ barely move, approaching to its nearest primary. 

The stability of the fixed points is shown in Table \ref{table_4}. It can be observed that for small values of the radiation parameter, $\beta \in (0,0.003]$, the librations $L_6$, $L_7$, and $L_8$ are linearly stable. Nevertheless, as $\beta$ grows the stability of the fixed points is modified and only $L_7$ and $L_8$ remain stable for $\beta \in$ $[0.004,0.999]$. Finally, when the gravitational force equates the radiation pressure force ($\beta = 1$), all the fixed points become unstable.

\begin{table}[t!]
\centering
\caption{Existence of equilibrium points with the variation of the radiation factor $\beta$, for the Sun-Jupiter-Trojan Asteroid system. \label{table_4}}
\vspace{0.2 cm}
{\begin{tabular}{cccc}
\hline
{\text{Interval}} & {\text{Equilibria}} & {\text{Stable equilibria}}\\ 
[0.1cm]\hline\\
[-0.2cm]
$\beta \in [0.000, 0.003]$ & $L_{1,2,3,4,5,6,7,8}$ & $L_{6,7,8}$\\
$\beta \in [0.004, 0.999]$ & $L_{1,2,3,5,7,8}$ & $L_{7,8}$\\
$\beta = 1.000$ & $L_{2,5}$ & $-$\\
\hline
\end{tabular}}
\end{table}

Taking into account that in the Solar system $\beta\approx 10^{-1}$, it is possible to conclude that in the Sun-Jupiter-Trojan Asteroid system, only the libration points $L_7$ and $L_8$ are stable, while the remaining equilibria are unstable $L_{1,2,3,5}$. Furthermore, as can be observed in Fig. \ref{fig_4} for $\beta=0$ the libration points $L_1$ and $L_2$ are symmetric about the position of $m_2$ on the $x-$axis, however for $\beta>0$ the symmetry is broken, with $L_1$ moving toward the Sun and $L_2$ toward Jupiter. In spite of this asymmetry, yet taking into account its similarities with the PR3BP, it would be interesting to consider possible applications to solar sail propulsion around $L_1$ and $L_2$, to provide station-keeping at quasi-periodic orbits around Jupiter \cite{BM08}.

\section{Orbit classification}
\label{results}

To investigate how the orbital structure of the system is affected by the radiation factor, let us start by defining the numerical criteria used for orbit classification. In general, there are three possible classes of trajectories for the fourth body: (i) escape to infinity, (ii) bounded motion around the primaries; (iii) collision with the primaries. The first two types of trajectories can be numerically determined as follows: Defining a disk of radius $R_e$, with center at the origin of the synodic frame of reference (center of mass of the Lagrange configuration of the PR4BP), the motion is considered as bounded if, after a certain integration time $t_{max}$, the fourth body stays confined inside the region delimited by the disk. Otherwise, an orbit is considered as an escaping orbit if the fourth body intersects the disk with velocity pointing outwards at a time $t_{e} < t_{max}$. Following Nagler and Zotos \cite{N04, N05, Z15a, Z15b}, we set $R_e = 10$ and $t_{max} = 10^4$, to ensure that the test body has enough time to escape. Concerning the third type of orbit, we define collision with a primary if the test particle crosses (pointing radially inward) a disk of radius $R_c = 10^{-4}$, with center at the position of the primary.

Furthermore, it should be pointed out that bounded orbits can be sub-classified into regular and chaotic orbits.  To do so, we shall use the Smaller Alignment Index (SALI) which is a mathematical tool introduced by Skokos for distinguishing between ordered and chaotic motion in conservative dynamical systems \cite{S01}. According to the numerical value of SALI at the end of the numerical integration, all bounded trajectories can be classified into regular (SALI $>10^{-4}$) or chaotic orbits (SALI $<10^{-8}$). To compute SALI, we consider the evolution of two orthonormal deviation vectors $\vec{w_1}$ and $\vec{w_2}$, which are continuously normalized to avoid overflow. Then, the smaller alignment
index is defined as SALI $\equiv \min(d_{-}, d_{+})$, with 

\begin{equation}
d_{\mp} \equiv \left\Vert \frac{\vec{w_1}}{ \left\Vert \vec{w_1}\right\Vert}\mp \frac{\vec{w_2}}{ \left\Vert \vec{w_2}\right\Vert}\right\Vert.
\end{equation}

In what follows, we take as a scattering region a $2 \times 2$ square grid, with a dense uniform grid of $625 \times 625$ initial conditions. Employing color-coded diagrams, we aim to show how the orbital structure is modified by the gradual variation of the radiation factor $\beta$ in the interval $[0,1]$. The color code is as follows: cyan $\rightarrow$ collision with $m_1$; red $\rightarrow$ collision with $m_2$; blue $\rightarrow$ collision with $m_3$; orange $\rightarrow$ escaping orbit; green $\rightarrow$ bounded regular orbit; yellow $\rightarrow$ bounded chaotic orbit. 

Lastly, it deserves mentioning that in case of collision and escape orbits, all calculations were performed using a double-precision adaptive Bulirsch-Stoer algorithm \cite{PTVF92}; while for the case of bounded orbits, since the deviation vectors need to be computed at the same integration time, a Prince Dormand Runge Kutta 8(7) \cite{PD81} was implemented. Also, for those orbits where the fourth body moves inside a region of radius $0.1$ around one of the primaries, the Levi-Civita regularization is applied to eliminate the singularities occurring in equations of motion (see e.g. \cite{BF02}). 

\subsection{Case 1: $m_1=m_2=m_3$}

\begin{figure*}[ht!]
\centering
\includegraphics[width = \linewidth]{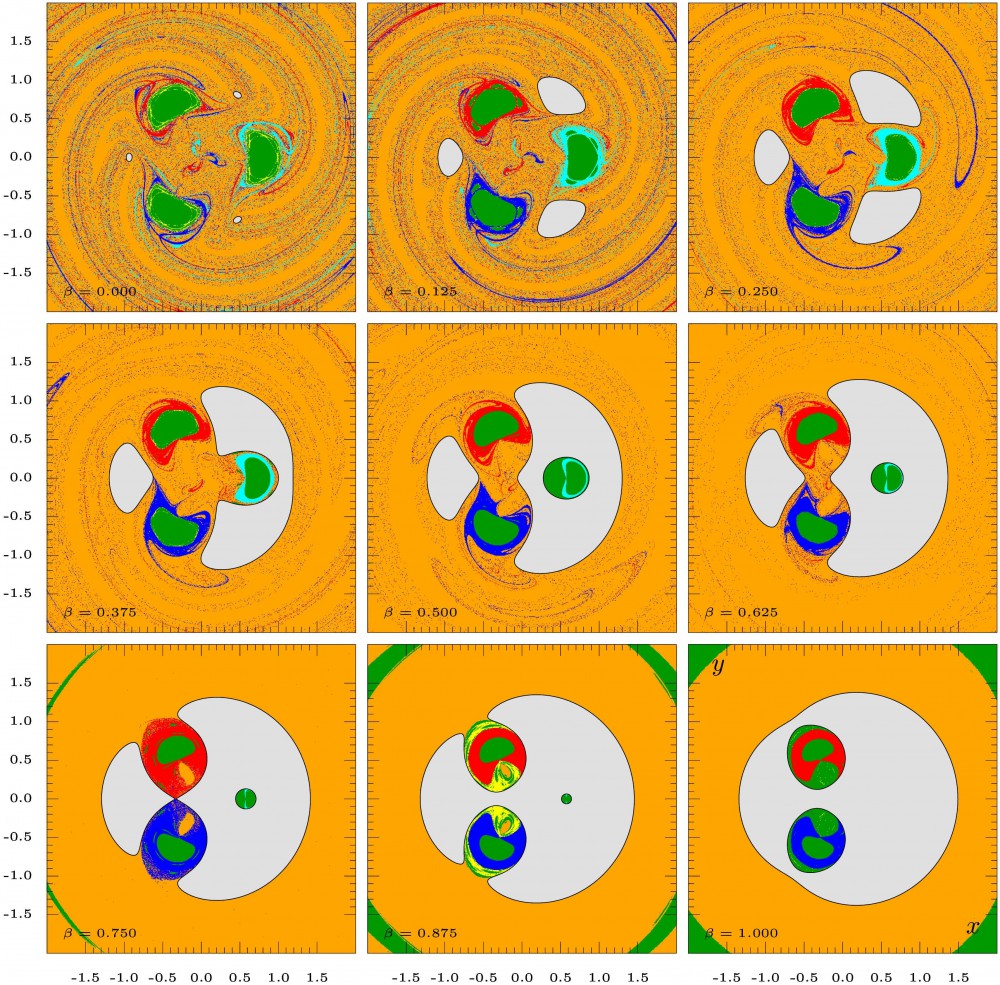}
\caption{(Color online). Orbital structure in the photogravitational Lagrange configuration of the PR4BP for the case $m_1=m_2=m_3=1/3$ when the radiation factor varies, setting $C=2.95$. The color code is as follows: cyan $\rightarrow$ collision with $m_1$; red $\rightarrow$ collision with $m_2$; blue $\rightarrow$ collision with $m_3$; orange $\rightarrow$ escaping orbit; green $\rightarrow$ bounded regular orbit; yellow $\rightarrow$ bounded chaotic orbit.} 
\label{fig_5}
\end{figure*} 

In this subsection, we will consider the case  $m_1=m_2=m_3$. Here, a characteristic value of the constant of motion which allows to clearly observe the evolution of the configuration space is $C=2.95$. In Fig. \ref{fig_5} we plot nine panels showing the evolution of the orbital structure for $\beta=0, 0.125, 0.25, 0.375, 0.5, 0.625, 0.75, 0.875$ and 1, respectively. 

Our numerical results suggest that in the absence of radiation $\beta=0$ (see the upper-left panel in Fig. \ref{fig_5}), the primaries are surrounded by regular islands which are slightly mixed with chaotic layers. The exterior region is predominantly occupied by a region of escaping orbits mixed with collisional orbits. In the center of mass of the system, the collisional orbits take the form of a ``three-blade propeller'' such that each blade contains collisional orbits with a specific primary. In presence of radiation $\beta=0.125$, the main differences are the following: the regions associated to collisional orbits which surround the stability islands about the primaries become wider, while the region of escaping orbits is reduced due to increase in the forbidden regions of motion. For larger values of the radiation parameter (e.g. $\beta=0.5$), the chaotic zones tend to disappear, however, for $\beta=0.875$ bounded chaotic orbits appear again around the collisional orbits surrounding $m_2$ and $m_3$. Finally, it is seen that for $\beta=1$ (see the lower-right panel in Fig. \ref{fig_5}), the mixture of escaping orbits with collisional orbits disappear in the outer region since the zero-velocity surface disconnects the primaries with the outer region. Also, it should be noted that the region of escaping orbits is embedded in a region of regular orbits.

In general, it is observed that as $\beta$ increases the forbidden region also increases its area, disconnecting $m_2$ and $m_3$ from the outer region. Moreover, the final state related to the collision with $m_1$ tends to disappear. In accordance with the results presented in subsection \ref{3.1}, the structure of the configuration space becomes smoother for larger values of $\beta$ because the number of fixed points is reduced.

\subsection{Case 2: $m_1 \neq m_2 = m_3$}

\begin{figure*}[ht!]
\centering
\includegraphics[width = \linewidth]{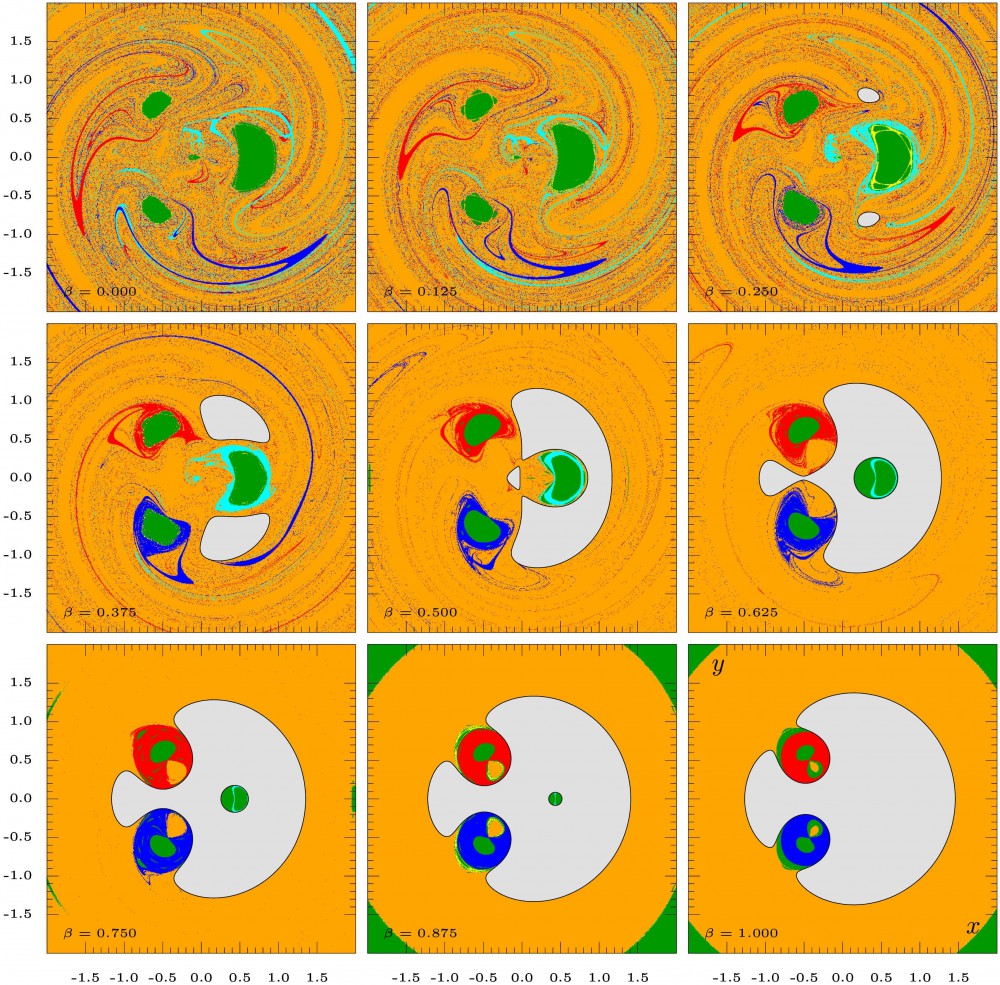}
\caption{(Color online). Orbital structure in the photogravitational Lagrange configuration of the PR4BP for the case $m_1= 0.5$, $m_2=m_3=0.25$ when the radiation factor varies, setting $C=2.7$. The color code is the same as in figure \ref{fig_5}.} 
\label{fig_6}
\end{figure*}

The second case corresponds to two equally massed primary bodies. As noted in subsection \ref{3.2}, the mass of the first primary is uniquely determined by ${\mathfrak{m}}$, hence, for the sake of simplicity, in what follows we assume $m_2=m_3=0.25$ and $m_1=0.5$. This selection is justified by the fact that (as can be seen in Fig. \ref{fig_3}) such configuration significantly differs from the first case ($m_1=m_2=m_3$) and the third case ($m_1\gg m_2, m_3$). Here, a numerical value of the constant of motion which allows observing the variation on the configuration space, when the radiation factor increases, is $C=2.7$. In Fig. \ref{fig_6} we plot the orbital structure in the configuration space for nine different values of the radiation factor $\beta$. The color-coded is the same used in the previous subsection.

The results for the present case can be summarized as follows: In absence of radiation $\beta=0$ (see the upper-left panel in Fig. \ref{fig_6}), each primary is embedded into a region of stability islands which are surrounded by a very narrow zone of chaotic orbits, however, unlike the previous case, these regions are not surrounded by collisional orbits. As in the case of equal masses, the outer region is mainly occupied by escaping orbits mixed with collisional orbits. In the presence of radiation $\beta=0.125$, the structures are very similar to the ones observed in the first panel, but the regions of regularity associated to $m_2$ and $m_3$ are subdivided forming an archipelago of 5 stability islands. 

For larger values of the radiation parameter (e.g. $\beta=0.5$), the chaotic zones tend to disappear and the number of regular islands around $m_1$ and $m_2$ is increased, being embedded into regions of collisional orbits. Also, it can be noted that the forbidden region isolates $m_1$, whose possible orbits are only collisional with $m_1$ or regular. When the radiation factor takes the value $\beta=0.875$, the outer region is formed by a large escaping region surrounded by a regularity zone. Here, chaotic zones appear in the inner and outer boundaries of the collisional regions of $m_2$ and $m_3$. Ultimately, for $\beta=1$ (lower-right panel in Fig. \ref{fig_6}), the foremost difference with the previous panels is that the boundaries between each kind of orbit are well established and the fractal-like structures disappear.

\subsection{Case 3: $m_1 \neq m_2 \neq m_3$}

\begin{figure*}[ht!]
\centering
\includegraphics[width = \linewidth]{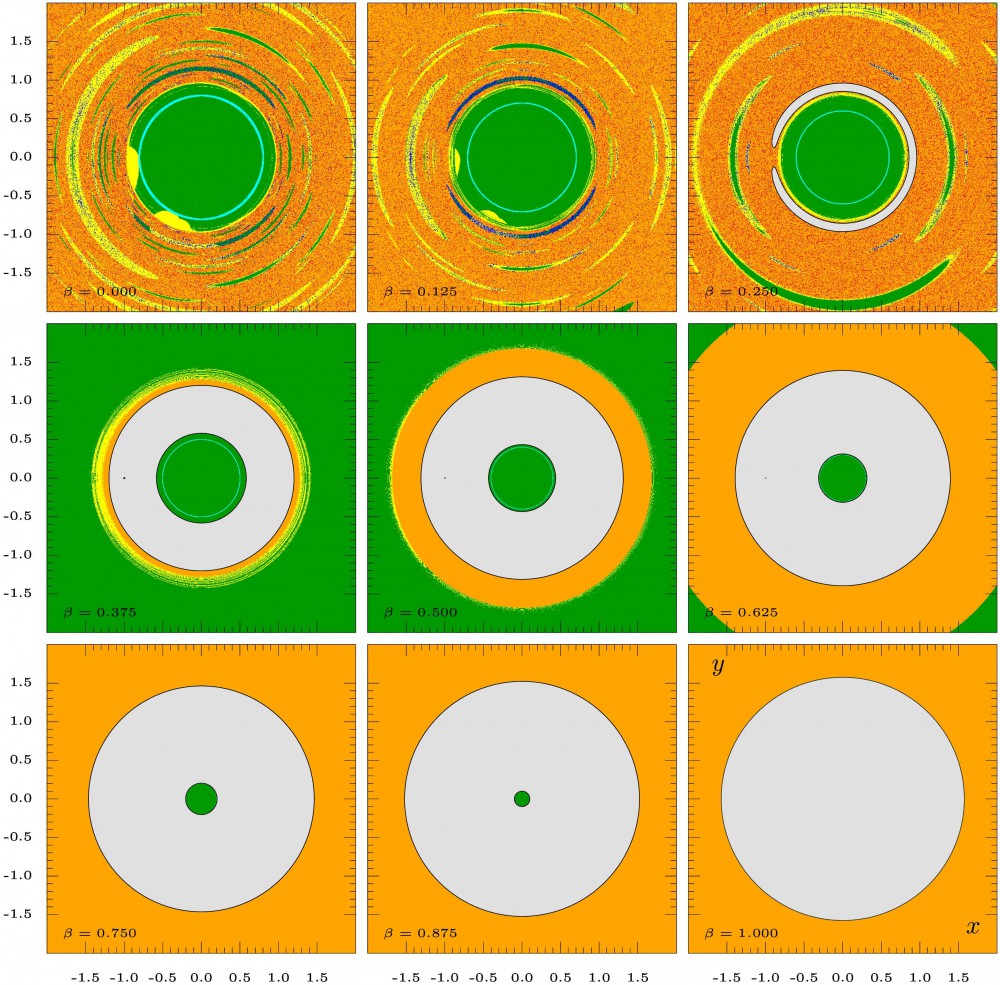}
\caption{(Color online). Orbital structure in the photogravitational Lagrange configuration of the PR4BP for the Sun-Jupiter-Trojan Asteroid system when the radiation factor varies, setting $C=2.485$. The color code is the same as in figure \ref{fig_5}.}
\label{fig_7}
\end{figure*}

The last case under consideration concerns to the configuration of different masses for the primaries. Following the reasoning of subsection \ref{3.3}, we will use the values of masses for the system Sun-Jupiter-Hektor-test particle, i.e., $m_1 = 0.999046321943$, $m_2 = 0.000953678050$, and $m_3 =  6.99996 \times 10^{-12}$.  It should be noted that in this case, the Sun is almost located at the origin of the coordinate system, Jupiter is approximately placed along the negative $x-$axis, and Hektor lies on the third quadrant of the coordinate plane.

With the increasing value of the radiation parameter $\beta$, the most remarkable changes in the orbital structure of the $(x, y)$ plane are shown in Fig. \ref{fig_7}. When the radiation is absent $\beta=0$, a circular region of stability surrounds $m_1$, with the appearance of an inner well-defined circle of collisional orbits with $m_1$. The positions of the primaries $m_2$ and $m_3$ are enclosed into a region of chaotic orbits, while the outer region is composed by a mixture of chaotic, regular, collisional and escaping orbits. For $\beta \ge 0.25$, the forbidden region of motion close itself off and hence the orbital structure drastically changes. 

The main changes are related to the displacement of the chaotic region to the exterior border of the forbidden region and the increase of the extension in the region of stability islands. The tendency for $\beta>0.625$ is to reduce the extension of the region of regular orbits and to increase the region of escaping orbits (see the last row of panels in Fig. \ref{fig_7}). 
 
\subsection{A global overview} 

\begin{figure*}[ht!]
\centering
\includegraphics[width = \linewidth]{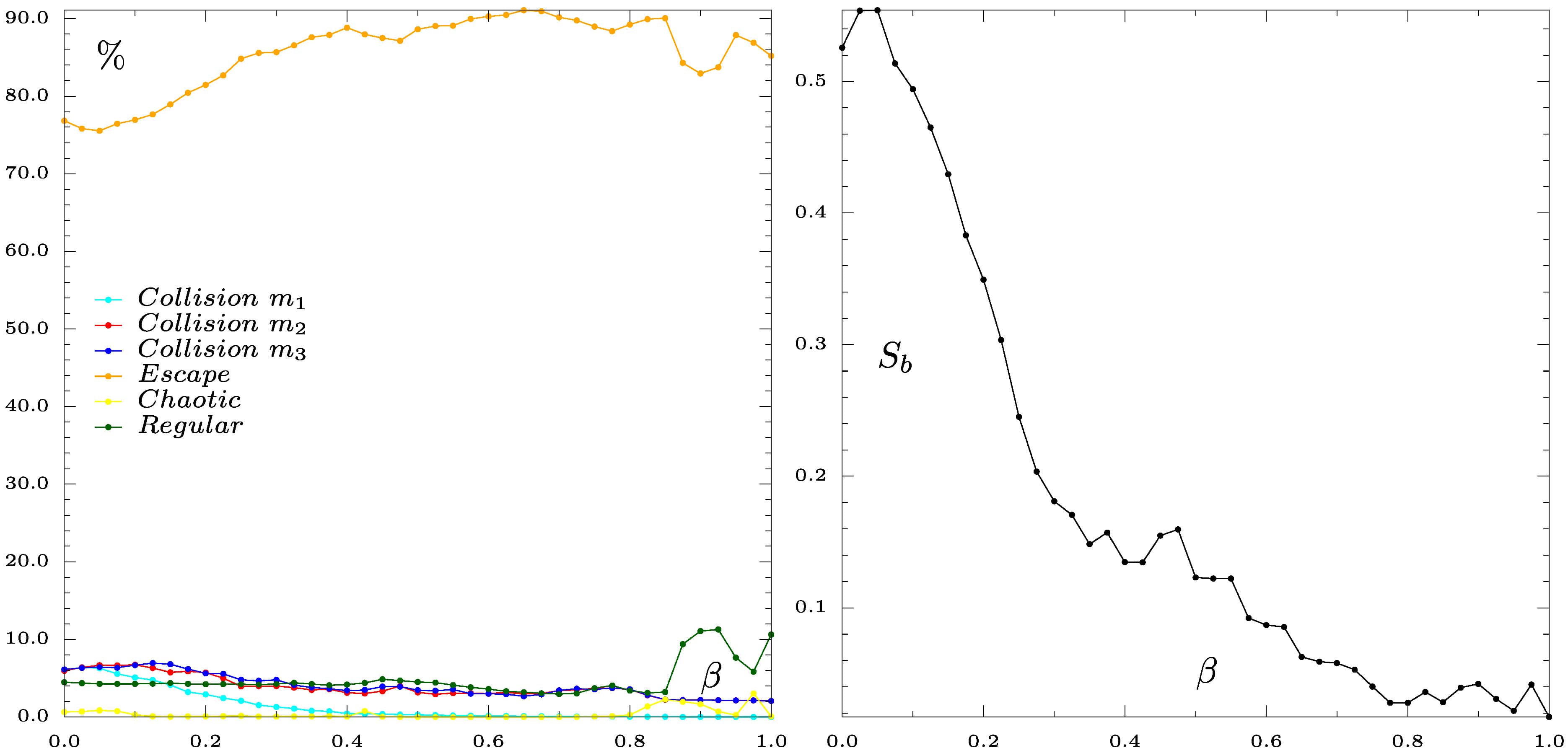}
\caption{(Color online). Parametric evolution of the percentages of each kind of orbit (left) and basin entropy $S_b$ on the $(x,y)$ plane (right), as a function of the radiation parameter $\beta$.  All other parameters have been set as in Fig. \ref{fig_5}.}
\label{fig_8}
\end{figure*}

Hereabouts, we present a detailed overview of the orbital dynamics in the photogravitational Lagrange configuration of the PR4BP. Each case studied in the previous subsections will be monitored by observing the evolution of the percentage of all types of orbits as a function of the radiation parameter $\beta$. As indicated at the beginning of section \ref{results}, in general, we have 6 classes of orbits: collision with $m_1$, collision with $m_2$, collision with $m_3$, escape, chaotic and regular. Therefore, the percentage of orbits gives us a quantitative measure of the number of orbits of each type present in the configuration space, or in other words, the most likely final state of a given initial condition.

Moreover, since our figures related to the orbital structure are in essence a set of initial conditions that are classified according to the final state of the trajectories, these figures can be understood as a sort of basins of convergence. Hence, as quantitative an indicator of the intricate distribution of initial conditions corresponding to each orbit, we also plot the basin entropy introduced by Daza et. al \cite{DWGGS16} which is a new tool to analyze uncertainty in dynamical systems. The basin entropy is defined as
\begin{equation}
S_{b}=\frac{1}{N}\sum_{i=1}^{N}\sum_{j=1}^{N_A}P_{i,j}\log\left(\frac{1}{P_{i,j}}\right).
\end{equation}
where $P_{i,j}$ denotes the the probability that inside a box $i$ the resulting final state is $j$, $N_A$ is the total number of final states and $N$ the total number of cells in which the whole region is subdivided\footnote{For a detailed explanation of the method, we refer the interested reader to Refs. \cite{DWGGS16} and \cite{DGGWS17}.}.

In Fig. \ref{fig_8}, it is observed that for the case of equal masses ($m_1=m_2=m_3=1/3$) the majority of orbits occupying the configuration space corresponds to escaping orbits, and the minority percentage for $0<\beta<0.8$ corresponds to chaotic orbits. For $0.8 < \beta \le 1$, the percentage of collisions with $m_1$ tends to zero, while the regular orbits stand in the second position oscillating between 4\% and 12\%. The basin entropy shows a gradual tendency to be reduced, which is in line with the results presented in subsection \ref{3.1}, suggesting that the uncertainty of the basins is larger for the classical Lagrange configuration of the PR4BP than for the photogravitational system with $\beta>0.3$.

\begin{figure*}[ht!]
\centering
\includegraphics[width = \linewidth]{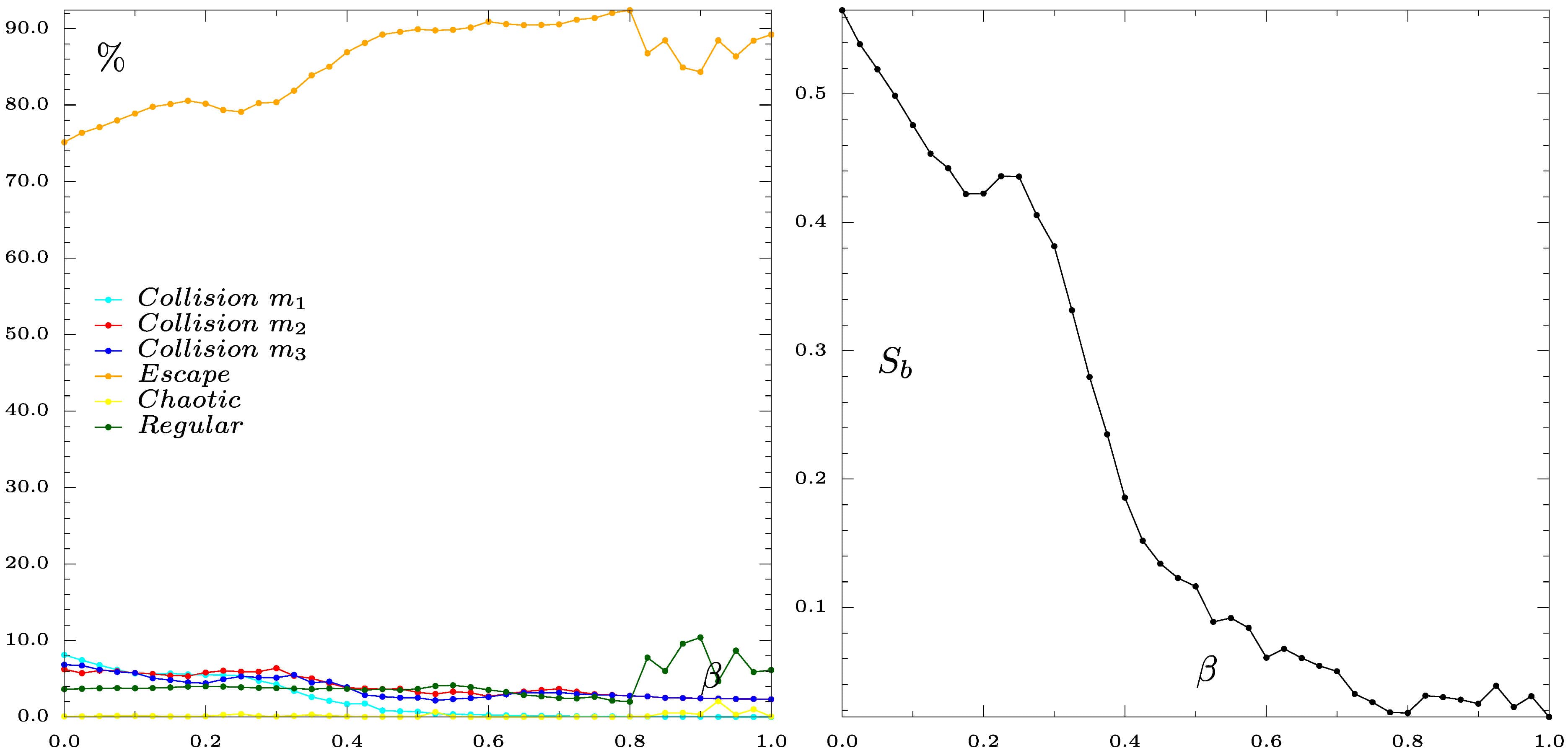}
\caption{(Color online). Parametric evolution of the percentage of each kind of orbit (left) and basin entropy $S_b$ on the $(x,y)$ plane (right), as a function of the radiation parameter $\beta$. All other parameters have been set as in Fig. \ref{fig_6}.}
\label{fig_9}
\end{figure*}

In Fig. \ref{fig_9}, we present our findings for the case of two equal masses ($m_1=0.5$, $m_2=m_3=0.25$). The results are very similar to the ones observed in Fig. \ref{fig_8}, i.e., the vast majority of orbits occupying the configuration space are escaping orbits while chaotic orbits are the minority percentage, except for $0.8 < \beta \le 1$, where the percentage of collisions with $m_1$ tends to zero. In the same way, the basin entropy exhibits a global tendency to be reduced up to the final value $S_b=0.02$.

\begin{figure*}[ht!]
\centering
\includegraphics[width = \linewidth]{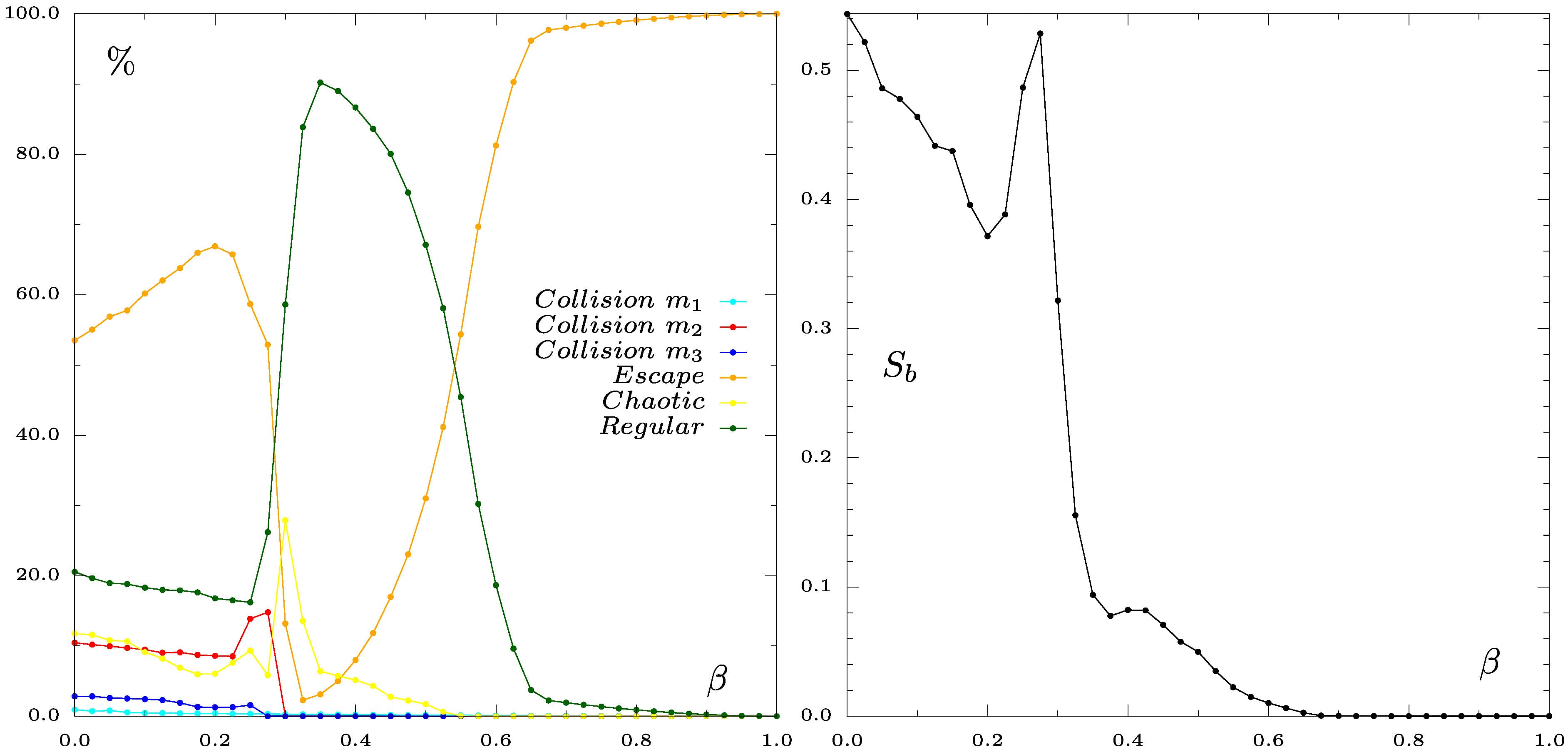}
\caption{(Color online). Parametric evolution of the percentage of each kind of orbit (left) and basin entropy $S_b$ on the $(x,y)$ plane (right), as a function of the radiation parameter $\beta$. All other parameters have been set as in Fig. \ref{fig_7}.}
\label{fig_10}
\end{figure*}

Finally, in Fig. \ref{fig_10} it is observed that in the case of different masses ($m_1=m_S$, $m_2=m_J$, $m_3=m_A$), the trend of the previous cases is broken. Although the number of escape orbits still being dominant for $\beta<0.3$ and $\beta>0.6$, in the interval of 0.3 up to approximately 0.6, the percentage is dominated by regular orbits. Also, it should be noted that in the interval $0\le \beta<0.2$, the percentage of regular orbits is about 20\% followed in number by chaotic orbits and collisions with $m_2$. It should be pointed out that approximately around $\beta=0.55$ the percentage of escape and regular orbits are equally distributed. From this value onwards, the increment in the percentage of escape orbits is almost the same as the reduction in the percentage of regular orbits. The breaking of the trend about $\beta=0.3$ is also present in the basin entropy. In the right panel of Fig. \ref{fig_10}, it can be observed that for $0<\beta<0.2$ the basin entropy tends to decrease, however for $0.2<\beta<0.3$ the tendency is inverted with a growing entropy, returning to a constant reduction up to approximately zero for values of the radiation parameter larger than 0.8.

\section{Concluding remarks}
\label{conc}

We numerically explored the orbital dynamics in the Lagrange configuration of the photogravitational planar restricted four-body problem with a unique radiating body. In particular, we demonstrated how the radiation parameter affects the position as well as the linear stability of the libration points in all possible combinations of mass for the primaries: equal masses, two equal masses, and three different masses.

Additionally, we performed a systematic analysis of the orbital structure on the configuration space by monitoring the evolution of the percentage of all types of orbits as a function of the radiation parameter. In general, we distinguish the following types of orbits: collision with each primary, escape, chaotic and regular. Another important aspect of this work was the calculation of the basin entropy to quantitative measure the uncertainty of the basins of convergence formed by the set of final states of the orbits.

The most important conclusions of our study are the following:

\begin{itemize}
\item In general, we observed that for larger values of the radiation parameter, the total number of fixed points tends to be diminished.
\item When the primary bodies are equal mass, all the equilibrium points are unstable regardless of the value of the radiation parameter.
\item In case of two equal masses for the primaries, the libration points $L_2, L_3, L_4, L_7, L_8, L_9,$ and $L_{10}$, are always unstable no matter the values set for the radiation factor or the mass for the primaries.
\item In the configuration of the Sun-Jupiter-Trojan Asteroid equilateral triangle, only the $L_7$ and $L_8$ libration points are stable, which are roughly located at the position of the non-collinear equilibrium points of the planar restricted three-body problem formed by the primaries Sun-Jupiter.
\item As a consequence of the reduction in the total number of fixed points in the system for larger values of radiation parameter, the basin structure of the configuration space becomes smoother.
\item Independent of the value of radiation for the first primary, the cases of equal masses and two equal masses are dominated, to a large extent, by the presence of escaping orbits. 
\item The parametric evolution of the percentage of orbits in the Sun-Jupiter-Trojan Asteroid , shows that for intermediate values of the radiation parameter (c.a. $[0.3, 0.55]$), the vast majority of orbits occupying the configuration space are regular orbits. 
\item The basin entropy in the Sun-Jupiter-Trojan Asteroid reaches a local maximum (similar to the one of the non-radiating case) for values of the radiation factor about $\beta=0.3$
\end{itemize}

We hope that the current numerical results to be useful in the active field of the dynamics of the four-body problem. Taking into account that our present outcomes are inspiring, as well as positive, it is in our plans to extend our investigation to two and three radiating primaries with different combinations of mass.

\section*{Acknowledgments}
\footnotesize
FLD acknowledges partial support from Universidad de los Llanos. FLD and GAG gratefully acknowledge the financial support provided by COLCIENCIAS (Colombia), Grant No. 8863. We thank the anonymous reviewers for their careful reading of our manuscript and their many insightful comments and suggestions.

\section*{References}

\interlinepenalty=10000

\end{document}